\documentclass{article}
\usepackage[utf8]{inputenc}
\usepackage{graphicx}
\usepackage[top=1in,left=1in,right=1in,bottom=1in]{geometry}
\usepackage{amsmath}
\usepackage{color}
\usepackage{dcolumn}
\usepackage{amsmath}
\usepackage{amsfonts}
\usepackage{amssymb}
\usepackage{bm}
\usepackage{epstopdf}
\usepackage{picture}
\usepackage{enumitem}
\usepackage{afterpage}
\usepackage{placeins}
\usepackage{float}
\usepackage{mathdesign}
\usepackage[english]{babel}
\usepackage{xr}
\graphicspath{ {./Figs/} }

\makeatletter
\newcommand*{\addFileDependency}[1]{
  \typeout{(#1)}
  \@addtofilelist{#1}
  \IfFileExists{#1}{}{\typeout{No file #1.}}
}
\makeatother

\newcommand*{\myexternaldocument}[1]{
    \externaldocument{#1}
    \addFileDependency{#1.tex}
    \addFileDependency{#1.aux}
}

\myexternaldocument{supplinfo}

\begin{document}
\title{Observation of Light-Driven Levitation Near Epsilon-Near-Zero Surfaces}
\author{M. G. Donato$^{1,*,\dagger}$, M. Hinczewski$^{2, }$\footnote{These authors contributed equally to this work.}\ , T. Letsou$^{2,3}$, M. ElKabbash$^{4}$, R. Saija$^{1,5}$,\\ P. G. Gucciardi$^{1}$, N. Engheta$^{6,\dagger}$, G. Strangi$^{2,7,\dagger}$, O. M. Marag\`{o}$^{1, }$\footnote{Email: maria.donato@cnr.it, engheta@ee.upenn.edu, gxs284@case.edu, onofrio.marago@cnr.it}}

\date{}

\maketitle

\noindent $^1$CNR-IPCF, Istituto per i Processi Chimico-Fisici, Consiglio Nazionale delle Ricerche, I-98158 Messina, Italy.\\
$^2$Department of Physics, Case Western Reserve University, 10600 Euclid Avenue, Cleveland, OH 44106, USA.\\
$^3$Harvard John A. Paulson School of Engineering and Applied Sciences, Harvard University, Cambridge, MA, 02138, USA\\
$^4$James C. Wyant College of Optical Sciences, University of Arizona, Tucson, Arizona 85721, USA. \\
$^5$Dipartimento di Scienze Matematiche e Informatiche, Scienze Fisiche e Scienze della Terra, Università di Messina, I-98166 Messina, Italy.\\
$^6$Department of Electrical and Systems Engineering, University of Pennsylvania, Philadelphia, PA 19104, USA.\\
$^7$NLHT Labs - Department of Physics and CNR-NANOTEC Istituto di Nanotecnologia, University of Calabria, Via Pietro Bucci 87036, Rende, Italy.\\

%\textbf{Abstract} 
{\bf Optical manipulation of micro- and nanoparticles near surfaces is fundamental for applications in sensing and microfluidics, yet controlling particle-surface interactions remains challenging. Here we experimentally investigate light-induced forces on dielectric particles near epsilon-near-zero (ENZ) metamaterial surfaces using photonic force microscopy. By illuminating trapped particles with tunable visible light, we observe a wavelength-dependent repulsive force unique to ENZ surfaces, contrasting with the attractive forces near dielectric or metallic substrates. This repulsion peaks near the ENZ frequency and may be attributed to combined optical ENZ effects and thermophoretic forces. Our findings demonstrate that ENZ metamaterials can induce stable levitation of particles via light-driven forces, offering a novel mechanism for contactless manipulation in microfluidic environments. This work advances understanding of light-matter interactions at ENZ interfaces and suggests potential for ENZ-based optical control of micro- and nanoscale objects, with potential applications in micro- and nanofluidic environments.} \\

%\section*{Introduction}
Small particles can be optically trapped at the micro and nanoscale \cite{Ashkin1970,Ashkin1986,Marago2013,juan2011plasmon,toftul2024radiation} and used as a very sensitive probe of the interaction with a sample, obtaining an all-optical scanning-force microscope \cite{GhislainOL93}, the \textit{photonic force microscope}, which has been proven ideal to test soft structures in liquid environments \cite{FlorinJSTRBIO97, bartsch2016nanoscopic, gao2025harnessing}. 
These measurements typically occur in the proximity of surfaces \cite{SchaefferLANGM07} necessitating the consideration of interaction forces between particles and the surface. Short-range repulsive or attractive forces, stemming from electrostatic and Van der Waals interactions \cite{Ruths2013}, Casimir effect \cite{sukenik1993measurement,munday2008measurements,munday2009measured,le2018measurement,pires2021probing,callegari2021optical}, or particle hydrodynamics influenced by nearby walls, must be accounted for in the total force evaluation  \cite{SchaefferLANGM07,Liu2014absolute, Shilkin2016near}. 
Consequently, precise calibration of the trap is essential for accurately measuring dynamic axial forces and surface force gradients  \cite{SchaefferLANGM07, Tolic-Norrelike, volpe2006surface, donato2021improved}. 

Furthermore, the unique characteristics of the surface itself may introduce additional forces that require consideration. For instance, when particles are positioned in front of Ag or Au surfaces, excitation of plasmonic resonances leads to light absorption and subsequent heat generation, resulting in convective and thermophoretic forces \cite{Baffou2017Thermoplasmonics, Kang2015plasmonic, Chen2015optofluidic, schmidt2025three}. Moreover, epsilon-near-zero (ENZ) metamaterials are associated with intriguing phenomena, such as theoretically predicted repulsive forces akin to the Meissner effect in superconductors \cite{Rodriguez2014, Engheta2013pursuing}. The Meissner effect provides a paradigmatic case where field expulsion at a boundary creates a repulsive potential capable of levitating nearby matter \cite{kim2019superconductor}. At certain frequencies, the real part of the permittivity in ENZ metamaterials changes sign, leading to the theoretical proposal and experimental observation of various exotic properties \cite{Liberal2017near, xie2025resonant, naserpour2025harnessing}. 

\begin{figure}
\centering
\includegraphics[width=\textwidth]{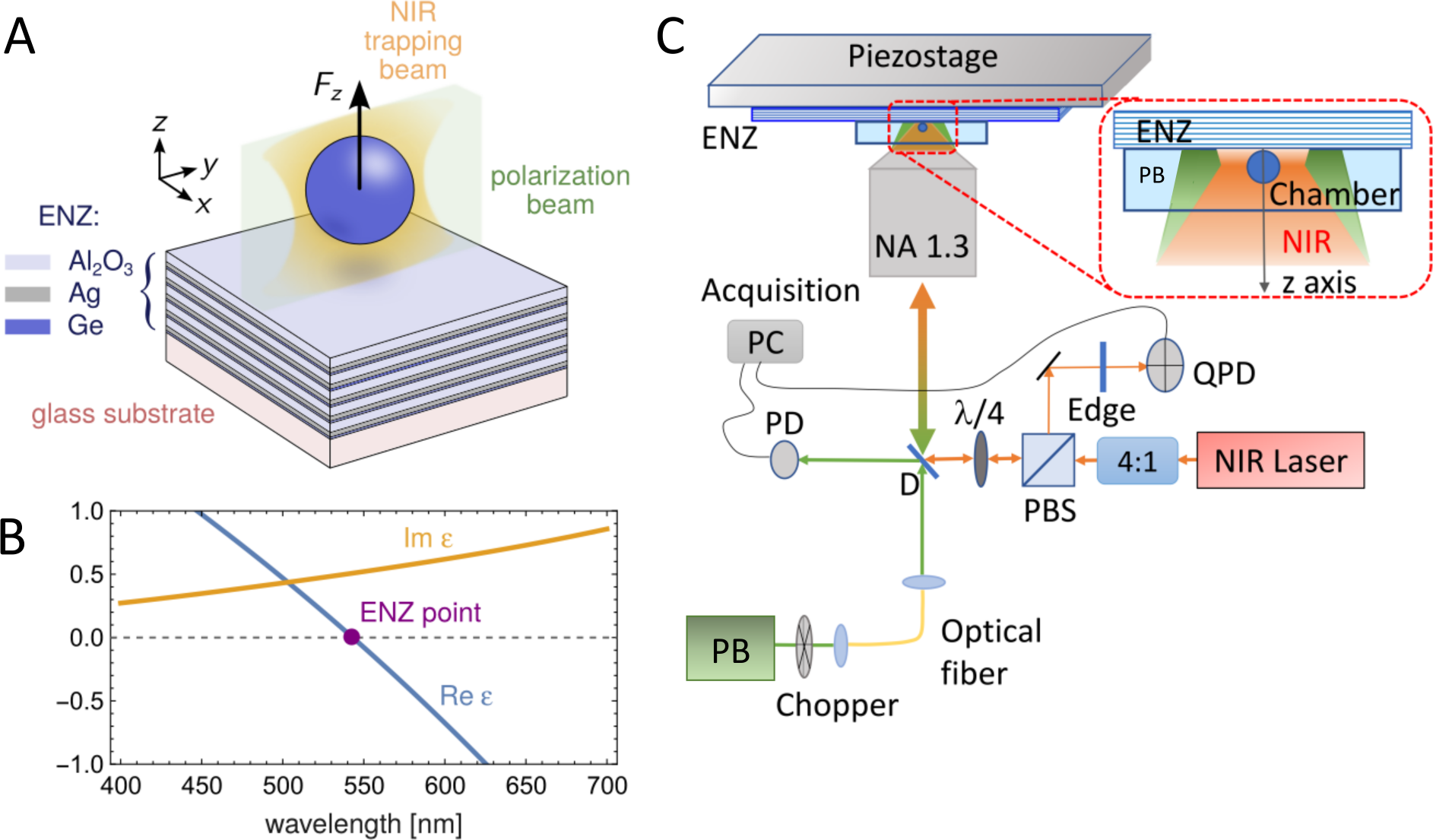}
\caption{\textbf{Epsilon-near-zero sample structure and experimental setup.} (\textbf{A}) A typical epsilon-near-zero (ENZ) sample structure consists of 5 trilayers of Al$_2$O$_3$ and Ag with a thin Ge layer ensuring surface wetting (see Methods). The stacks are deposited on a glass substrate. A polystyrene probe particle is trapped near the surface and excited with tunable visible light to measure the resulting axial force. (\textbf{B}) Optical properties of the ENZ sample structure. The multilayer is deposited by electron beam evaporation and designed to exhibit an ENZ frequency with low loss, with a nominal complex permittivity $\epsilon_s = 0.51 i$. The real and imaginary parts of the effective permittivity are obtained from spectroscopic ellipsometry combined with transfer-matrix modeling, showing that the ENZ condition, ${\rm Re} (\epsilon_s) = 0$, occurs at a wavelength of about 547 nm. (\textbf{C}) Sketch of the experimental setup. A laser diode provides the trapping beam at 830 nm. After passing through a 4:1 telescope, a $p$-polarizing beam splitter (PBS) and a quarter waveplate, it is reflected by a dichroic mirror (D) towards the back aperture of a NA=1.3 objective, which focuses the trapping beam to the diffraction limit near the surface inside the sample chamber (see inset, not to scale). A piezostage with nanometric resolution is used for positioning and motion control. The backscattering from the trapped particle is back-collected by the same objective, passes through the quarter waveplate and its $s$-polarized component is delivered to the detector, a quadrant photodiode (QPD). A long-pass edge filter prevents light at wavelengths $<$ 633 nm to reach the detector. A white light laser is used as a tunable polarizing beam (PB) by selecting its wavelength in the visible. The light is sent through a fiber and focused so that the beam waist at the sample is $w_0=2.8\pm0.2$ $\mu$m (larger than the particle radius). The PB is chopped at 8 Hz and sent through an optical fibre on the trapped particle. A photodiode (PD), collecting a small portion of the PB, is used to synchronize the tracking signal measurements with the laser pulses. The signals from QPD and photodiode are acquired by a data acquisition board and stored in a computer for the analysis.}
\label{Fig1}
\end{figure}

In this regime, the continuity of the displacement field forces electromagnetic fields to bend away from the interface, forming an optical potential whose repulsive character parallels the magnetic-field expulsion observed in superconductors.
Several theoretical studies have suggested that this ENZ-induced expulsion of displacement field lines should generate a measurable photonic analogue of Meissner-type levitation  \cite{Rodriguez2014,Engheta2013pursuing,Liberal2017near}. Building upon our previous work \cite{kiasat2023epsilon}, which theoretically explored the optomechanics of nano- and microscale particles in the vicinity of epsilon-near-zero (ENZ) and other surfaces, with a repulsive-attractive behavior depending on surface characteristics and particle polarizability, in the present work we experimentally investigate this phenomenon. Despite the robustness of these predictions, no experiment has yet directly measured the near-field optical forces that an ENZ interface is expected to exert on a nearby particle. Previous studies have focused principally on ENZ-enhanced nonlinearities, field uniformity, and phase control, leaving the force–distance landscape almost entirely unexplored \cite{wang2016strong, shi2022optical}. In this study, as shown in Fig. \ref{Fig1}, we utilize photonic force microscopy to quantify forces exerted on a particle in proximity of surfaces with varying compositions and reflective properties. The probe used is a dielectric polystyrene particle of 1 $\mu$m radius, trapped in water by a near-infrared (NIR) laser beam (Fig. \ref{Fig1}A,C). The trap is calibrated both in bulk (away from the surface) and close to the surface. After calibration, the position of the trapped particle is perturbed in the axial direction by a chopped light from a second tunable laser source, namely a laser beam at visible wavelengths (480 nm-620 nm range) that polarizes the particle. Under this illumination, an axial shift of the particle equilibrium position is observed, which can be used to measure not only the strength of the total axial force on the particle but also its attractive or repulsive character, depending on the intrinsic response of the different sample surfaces. In this way, we aim to achieve a comprehensive understanding of the forces governing particle-surface interactions. We compare and contrast our measured forces in front of dielectric-only, metallic-only and the ENZ metamaterials, providing insight into potential sources of levitating forces in the presence of ENZ scenario. For our dielectric-only surface, we use glass; for our metallic-only surface, we use Ag; and for our ENZ surface, we use a layered structure made of Al$_2$O$_3$, Ag, and Ge (see Fig. \ref{Fig1}A and Methods summary).  The effective dielectric function of the ENZ surface varies with wavelength, as shown in Fig. \ref{Fig1}B.  We acknowledge the significance of unavoidable long-range thermal effects, including thermophoresis and thermoconvection, in influencing particle dynamics near such surfaces. 

\subsection*{Results and discussions}
\begin{figure}
\begin{center}
\includegraphics[width=\textwidth]{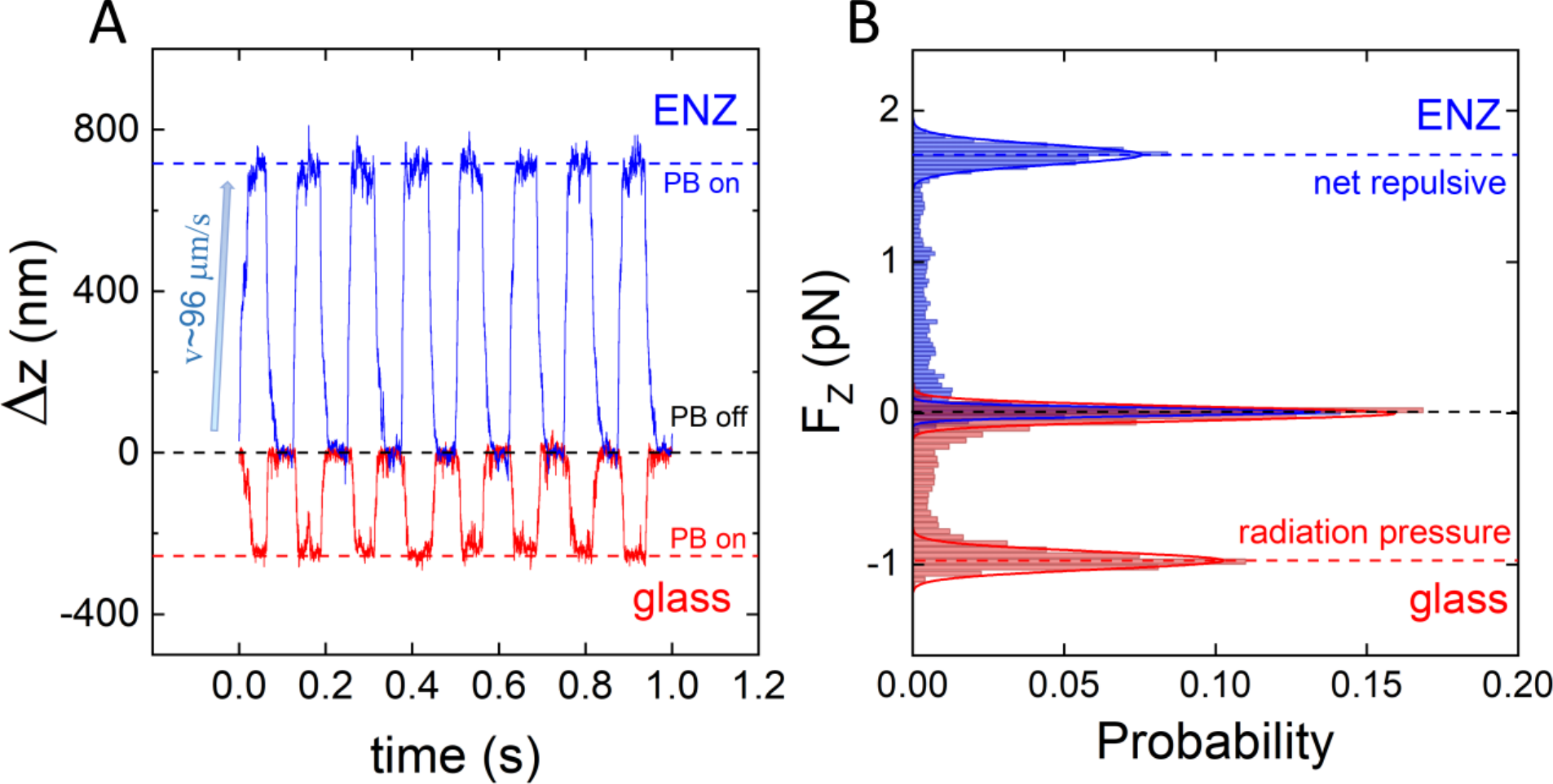}
\caption{\textbf{Calibrated optical force measurements near sample surfaces.} (\textbf{A}) Calibrated axial tracking signals under pulsed polarizing light from the PB source. The polystyrene particle is trapped in front of glass (red curve, $P_{trap}\sim $13 mW) or in front of ENZ (blue curve, $P_{trap}\sim $2 mW) surface. The polarizing white light beam is chopped with a frequency of 8 Hz. The trap equilibrium positions when PB illumination is on (red dashed line, glass, $\lambda_{\rm PB}=620$ nm, $P=2.6$ mW; blue dashed line, ENZ, $\lambda_{\rm PB}=500$ nm, $P=1.77$ mW) or off (black dashed line) are shown. Negative/positive shifts correspond to axial motion towards/away from the surface. The inset shows the average speed of the particle repelled by the ENZ surface, $\sim 96\ \mu$m/s (see Suppl. Mater.).  (\textbf{B}) Axial force distributions (red, glass surface; blue, ENZ surface) corresponding to the tracking signals in (A). These are obtained from the optical trap calibration through the power spectrum density analysis of the particle thermal fluctuations (see Suppl. Mater.).}
\label{Fig2}
\end{center}
\end{figure}

%\paragraph{Force measurements with chopped light.}%
As mentioned earlier, to polarize trapped particles, we used a second laser beam with wavelength selected in the range 480 nm-620 nm (step 20 nm and bandwidth 10 nm) from a white light polarizing beam (PB) source (see Supplementary Materials). Square-wave pulses at 8 Hz have been produced by a chopper in the PB path. PB pulses affect the particle equilibrium position in the trap \cite{zensen2016pushing} and, thus, the backscattered NIR signal. In Fig. \ref{Fig2}A, the signals obtained in front of glass (red curve) and ENZ (blue curve) are shown. After suitable calibration in the axial direction (see Suppl. Mater.), the shift $\Delta z$ of the trapped bead due to polarizing beam pulses can be calculated (Fig. \ref{Fig2}A). Particle velocity may reach values as high as 96 $\mu$m/s (Fig. \ref{Fig2}A and Suppl. Mater.). It is clear that the effect of polarizing beam illumination on the trapped particle is different in front of the two surfaces because the signals shift in opposite directions. In front of glass we expect that only the radiation pressure due to PB pulses can change the trapped particle position, pushing it towards the surface. The resulting total force is therefore attractive and, in our reference system, is represented as negative (Fig. \ref{Fig2}B, red). Conversely, when positioned in front of the ENZ surface, an opposing scenario is observed, indicating a repulsive force that is positive (Fig. \ref{Fig2}B, blue). 

After determining the absolute surface position and the medium temperature using the drag force method (see Suppl. Mater.), we calibrate the axial total force, $F_z$ under polarizing beam pulses at 8 Hz by using the power spectral density (PSD) of the tracking signals, where the presence of a periodic force becomes readily apparent  (see the Suppl. Mater. for details).

\begin{figure}
\begin{center}
\includegraphics[width=\textwidth]{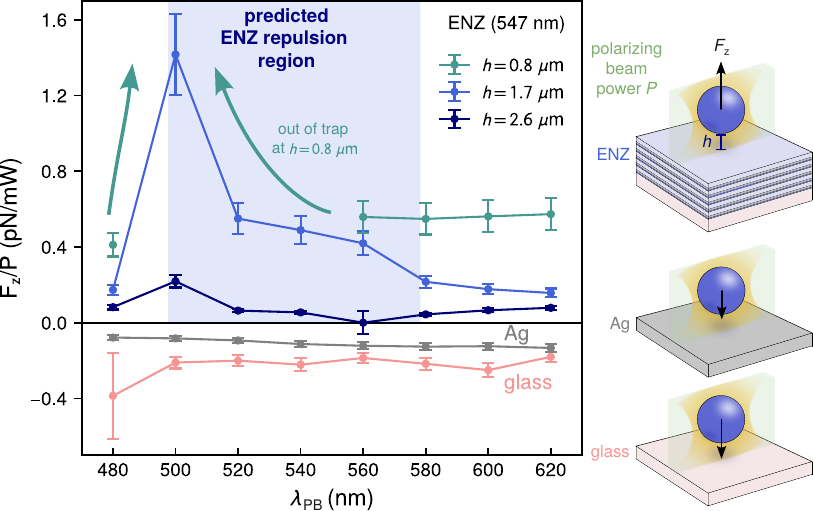}
\caption{\textbf{Total force versus wavelength for particles over different surfaces.} Total normalized axial force $F_z / P$ under pulsed polarizing beam illumination of incident power $P$ as a function of the polarizing beam wavelength $\lambda_{\rm PB}$.  The top three curves are for a particle over an ENZ at three different edge-to-edge heights $h$ above the surface, all showing net repulsion ($F_z > 0$). In contrast, the bottom two curves are over Ag (gray, {$h = 0.8$ $\mu$m}) and glass (pink, {$h= 8$ $\mu$m}), showing net attraction ($F_z < 0$). The three configurations are shown schematically on the right. The ENZ surface has an ENZ point at 547 nm, and the region shaded in blue shows the range of wavelengths where the optical force is predicted to be strongly repulsive. This range was approximately calculated as the full-width half-maximum of the $F_z > 0$ region from a simplified theory assuming a point dipole close to the surface, with the ENZ modeled via an anisotropic effective medium~\cite{rodriguez2016repulsion}. For the ENZ curves the force is dramatically enhanced in this region, and for $h= 0.8$ $\mu$m it becomes sufficiently strong to push the bead out of the trap, preventing force measurements between $\lambda_{\rm PB} = 500-540$ nm.}
\label{Fig3}
\end{center}
\end{figure}

The axial PSD used for the calibration of $F_z$ is the average over ten different PSDs, each one corresponding to 1 s observation time, at 50 kHz sampling rate. We acquired PSDs at several polarizing wavelengths $\lambda_{\textrm{PB}}$ and at different distances from dielectric, metallic or ENZ surfaces. The results obtained approaching ENZ surface and, as a comparison, in front of glass and Ag surface are shown in Fig. \ref{Fig3}A. 

%Fig.3

Forces exerted on the PS particle in front of glass or Ag surfaces (pink and gray points in Fig. \ref{Fig3}A, respectively) are always negative, indicating that the polarizing PB pulses push the particle towards the surface. Their magnitude is small and remains constant regardless of the wavelength. Conversely, forces in front of ENZ surfaces are repulsive, causing the particle to be pushed away from the surface. Furthermore, these forces exhibit a wavelength-dependent nature, with a maximum occurring around 500 nm, blue-shifted with respect to the nominal ENZ wavelength at 547 nm. If the same measurements are carried out in front of an another ENZ sample with ENZ wavelength at 590 nm, a similar blue shift of the maximum axial force is observed (Suppl. Mater.).

To understand further the origin of this repulsive force in front of the ENZ surface, we investigate the contributions from short-range forces, including electrostatic double layer, van der Waals forces, and Casimir forces. In case of electrostatic double layer and Van der Waals forces, the Debye-Landau theory elucidates the electrostatic repulsive potential stemming from the electrostatic double layer, alongside the attractive van der Waals potential, both contingent upon distance and the concentration of charge carriers in the liquid. However, it is noteworthy that these potentials are predicted to be influential only at distances on the order of tens of nanometers, as they diminish significantly before reaching particle-surface distances of 100 nm. Our measurements, indicating pN forces exerted on the dielectric bead, suggest that these short-range forces are unlikely to contribute significantly to the observed levitation effect.
Considering the interaction between the dielectric particle and the layered structure, particularly with materials like Al$_2$O$_3$ or TiO$_2$ as the first layer, the potential influence of Casimir forces is evaluated. Prior research notes that the Casimir force between two dielectric surfaces peaks at approximately 10 fN at edge-to-edge distances of 400-500 nm, diminishing before reaching distances of 1 $\mu$m. While our measurements register forces in the pN range, indicative of potential Casimir force contributions, it's essential to consider the context. Thermal fluctuations, often associated with Casimir forces, might induce sticking effects on surfaces, but this mechanism primarily explains attraction, not repulsion. Thus, it's reasonable to infer that Casimir effects are not significant contributors to the observed repulsive behavior in front of the ENZ surface.
%Long Range Forces and Inference on Levitation Effect
In contrast to short-range forces, our analysis suggests that the optical ENZ force plays a crucial role in the observed levitation, while other forces arising from the energy deposited in the system may also contribute. Their possible interplay is being considered in future studies as a route to engineering advanced optomechanical systems. A more careful analysis of the measurements performed under particle polarization conditions reveals that, when generating the oscillating field with incident powers in the few mW range (1–5 mW), Faxen’s law consistently indicates a local temperature increase of approximately 10 K in the region occupied by the particle on ENZ substrates. This corresponds to an effective thermal gradient on the order of 1 K/$\mu$m. Thermal effects therefore emerge as a non-negligible contribution in our experimental configuration, with Faxen’s law converging to an estimated temperature gradient of $\sim 1$ K/$\mu$m under a particle polarization density of about 0.1 mW/$\mu$m$^2$. This gradient leads to the generation of a dragging force on dielectric particles through thermophoresis, resulting in a thermophoretic speed of approximately 1 $\mu$m/s. Comparatively, thermoconvective forces, operating under low Reynolds' numbers, induce speeds almost two orders of magnitude lower for the same thermal gradient \cite{lee2020thermophoretic}. Utilizing the Stokes dragging law, we estimate the thermophoretic force exerted on our PS bead is on the order of 1 pN \cite{franzl2022hydrodynamic}.
Moreover, our ability to extract and measure the ENZ optical force was facilitated by a specific experiment leveraging the frequency dependence of the ENZ response (Fig. \ref{Fig3}). This experimental setup provided crucial insights into the behavior of the ENZ force within its frequency range, further validating its significance in our observed levitating effect. Thus, these observations suggest that the combined action of the ENZ force and a thermophoretic contribution may underlie the overall measured force, consistent with the observed nonlinear dependence of force on polarizing power (see Supplementary Materials). This reinforces the significance of these long-range forces in our experimental observations.
Furthermore, our data highlights a remarkable frequency dependence of the total optical force within the epsilon-near-zero region: the dramatic peak in force around 500 nm is much steeper than the predicted peak in absorbed power for the ENZ substrate (Fig. \ref{FigAbsorption}), suggesting thermal effects alone cannot fully account for the observed forces. This behavior further underscores the complexity of interactions in this regime and emphasizes the need for comprehensive analysis to elucidate the underlying physics. In summary, our experimental findings are consistent with the involvement of the optical ENZ force and thermophoretic effect in the observed levitating effect, underscoring the importance of considering both short and long-range forces in understanding complex phenomena at the nanoscale.

\subsection*{Conclusions}
In summary, we have directly measured light-driven repulsive forces acting on a trapped microparticle in the near field of an epsilon-near-zero surface. By probing the system with wavelength-tunable pulses and comparing the particle response above dielectric, metallic, and ENZ interfaces, we identified a force component that emerges only in the vicinity of the ENZ crossing. Its wavelength selectivity, appearance at sub-milliwatt powers, and distinct scaling with intensity cannot be reproduced by absorption-driven heating alone. This repulsive behavior may reflect the ENZ-induced redistribution of the displacement field at the interface.
At higher powers, a thermophoretic contribution becomes appreciable and adds to the repulsion, but its spectral and power dependence remain separable from the ENZ optical force. The coexistence of these two mechanisms is what produces the characteristic power-dependent crossover observed near the ENZ point.
Taken together, these measurements provide an experimental evidence of a Meissner-like optical response in a classical photonic platform: an ENZ interface that expels electric displacement fields and generates a repulsive force on nearby matter. This establishes ENZ materials as a pathway to engineer synthetic levitation phenomena, enabling contactless control of micro- and nanoscale objects with minimal optical power. The ability to sculpt force landscapes through photonic boundary conditions opens opportunities in optomechanical manipulation, microfluidic transport, and sensing schemes where selective repulsion, levitation, or force shielding are required.

\section*{Methods summary} 
Additional details of the materials and methods can be found in the Supplementary Materials.

\subsubsection*{Optical tweezers}
The experimental set-up is sketched in Fig. \ref{Fig1}C. The trapping beam is provided by a $\lambda$=830 nm laser diode (Thorlabs
DL8142-201, 150 mW maximum power). The beam passes through a couple of anamorphic prisms and an optical isolator (both not shown) to be circularized and to avoid back-reflected light reaching the diode, respectively. A 4:1 telescope is used to enlarge the beam spot. After this, the beam is first \textit{p}-polarized by a polarizing beam splitter (PBS) and successively it is circularly polarized by a quarter waveplate. After the reflection from a dichroic mirror, it is directed towards the back-aperture of a high numerical aperture objective (Olympus, Uplan FLN 100X, NA=1.3), in overfilling condition. The resulting beam focal spot is about 0.4 $\mu$m in diameter (diffraction limited spot). The light scattered by a trapped particle in front of the surface is collected in backscattering configuration by the same objective and, after the reflection from the dichroic mirror, it passes through the quarter wavelength. Then, the \textit{s}-polarized component selected by the PBS is reflected towards a second telescope (not shown) and delivered to a quadrant photodiode (QPD, Thorlabs PDQ80A). To avoid visible light reaching the detector, an edge filter (Semrock, Razor Edge LP02-633RU-25) is put in front of the QPD. Occasionally, to avoid saturation of the detector, especially when trapping in front of highly reflective surfaces, neutral density filters are also put in front of the QPD. Typical trapping power at the objective was approximately 15 mW in front of glass, while only 4 mW in front of Ag and approximately 2 mW in front of ENZ, to avoid surface deterioration by the highly focused laser beam.

Polarizing beam (PB) with wavelength ranging from 480 nm to 620 nm (step 20 nm and bandwidth 10 nm) is provided by a white light supercontinuum laser source (SuperK Extreme, NKT Photonics) equipped with a variable bandpass filter (SuperKVaria, NKT Photonics). A chopper (Thorlabs, MC1000A) is inserted in the PB path to give square wave laser pulses at 8 Hz. The PB is delivered to the experiment through an optical fiber and it is mostly directed to the objective back aperture, while a negligible fraction is sent to a photodiode working as a trigger to synchronize the particle tracking acquisition. The PB spot has a diameter of approximately 6 $\mu$m in the trapping region. We checked that this beam cannot trap particles, due to its poor focusing and low intensity. PB maximum power at the focal region is approximately 2 mW (Fig. \ref{WL}). Finally, the voltage signals from QPD and photodiode are acquired by a data acquisition board (National Instruments PCI-6035E) and stored in a computer for the analysis.

Polystyrene spheres of 1 $\mu$m radius  (Polysciences Polybead\textregistered \ Microspheres) are used as probe particles. The sample chamber is built directly on the sample surface, by dropping approximately 10 $\mu$L of dispersion, and sealing the chamber with a No.1 glass coverslip and nail polish (see the inset in Fig. \ref{Fig1}C). Following this procedure, a typical sample chamber with a thickness of about 25 $\mu$m is formed.
The position of the sample chamber is controlled by a multi-axis piezostage (Physik Instrumente, P-517.3 CL) with subnanometer resolution.

\subsubsection*{Preparation of ENZ substrates}
The sample with ENZ wavelength at 547 nm consisted of 5 trilayers of Al$_2$O$_3$ (40 nm) / Ag (8 nm) / Ge (1 nm) from top to bottom, with thicknesses indicated in parentheses.  The thin Ge layers were used to ensure surface wetting. The stacks were deposited on a glass substrate (Corning Inc.) with electron-beam evaporation for Ge (0.5 $\text{\AA}  / \text{s}$) and Al$_2$O$_3$ (0.3 $\text{\AA}  / \text{s}$), and thermal evaporation for Ag (0.5 $\text{\AA} / \text{s}$). This process produces smooth, conformal coatings with reproducible thickness and optical properties compatible with the trapping and force-mapping experiments. The materials used in the fabrication were purchased from Kurt J. Lesker.  To determine thicknesses and refractive indices of the layers in each stack, ellipsometric measurements of both samples in an air superstrate were carried out using a variable-angle, high-resolution spectroscopic ellipsometer (J. A. Woollam Co., Inc, V-VASE) for incident angles $45^\circ$, $50^\circ$, $55^\circ$ and wavelength range $300 - 1000$ nm.  Once the parameters of the stack were established via best-fits to the ellipsometric data, the ENZ wavelength with a water superstrate was determined via the effective medium theory described in Ref.~\cite{kiasat2023epsilon}.

\section*{Data Availability}
Data that support the findings of this study are available from the corresponding authors upon reasonable request.

\section*{Acknowledgements}
This work has been funded by European Union (NextGeneration EU), through the MUR-PNRR projects SAMOTHRACE (ECS00000022) and PE0000023-NQSTI, the PRIN2022 "EnantioSelex" (2022P9F79R), “FLASH-2D” (2022FWB2HE), "Cosmic Dust II" (2022S5A2N7), and carried out within the Project“Space It Up” funded by ASI and MUR– Contract n. 2024-5-E.0 - CUP n. I53D24000060005. We thank F. Patti for help in the early stage of this work.

\section*{Author contributions}
M.G.D. and O.M.M. performed the optical trapping experiments.
M.H., M.E, T.L. and G.S. designed and fabricated the surface samples. M.G.D. and M.H. elaborated the data. R.S., P.G.G., G.S., and O.M.M. procured the funding. M.G.D., M.H., N.E., G.S., and O.M.M. wrote the manuscript. N.E., G.S., and O.M.M. conceived and supervised the work. All the authors discussed and commented on the results and on the manuscript text.

\section*{Supplementary Materials}
Materials and Methods\\
Supplementary Text (Power spectrum density analysis of a trapped particle fluctuations; Hydrodynamics of a sphere close to a surface and NIR trap calibration; Pulsed force on a trapped particle and power spectral density; Light-driven repulsive force as a function of ENZ excitation power: nonlinear behaviour, thermophoresis, and particle velocity; Absorbed power spectrum; Additional ENZ sample with a titania-silver layered structure) \\
Figs. S1 to S9\\
Additional references \cite{Jones2015,Berg2004,Faxen1922widerstand,Brenner1961slow}

\newpage

\centerline{\huge{Supplementary Materials}}

%\section*{Supplementary text}
\subsection*{Materials and Methods}
We use photonic force microscopy to measure the total force acting on a polarizable dielectric particle trapped in front of dielectric, metallic and ENZ surfaces and subjected to visible pulsed laser illumination. We expect different particle dynamics, due to the intrinsic peculiarities of each surface.  
The rationale guiding our measurements is to observe if some ENZ-based force can perturb the dynamics of a bead trapped in front of an ENZ surface illuminated by light at its ENZ wavelength.
To this purpose, we first trap a 1 $\mu m$ radius polystyrene particle at fixed distance from the surface by means of a near-infrared (NIR) laser at 830 nm. Then, we collect the NIR backscattered light and calibrate the optical trap. Thus, when the particle is illuminated by light pulses with tunable wavelength (480 nm-620 nm range) in the axial direction (see Fig. \ref{Fig1}), \textit{i.e.}, perpendicularly to the surface, we accurately measure the total force on the trapped particle and compare the results in front of the dielectric, metallic, and ENZ surfaces. This discriminates between forces due to the radiation pressure of the polarizing laser (always present) and forces due to an intrinsic response of the different sample surfaces.
Furthermore, in the ENZ case, when illuminated by the polarizing laser light, the dynamics of the trapped particle is wavelength dependent, as the surface permittivity is changing dramatically across the explored wavelength range \cite{Rodriguez2014} 
(Fig. \ref{Fig1}\textbf{B}).

\begin{figure}[h]
\begin{center}
\includegraphics[width=0.5\textwidth]{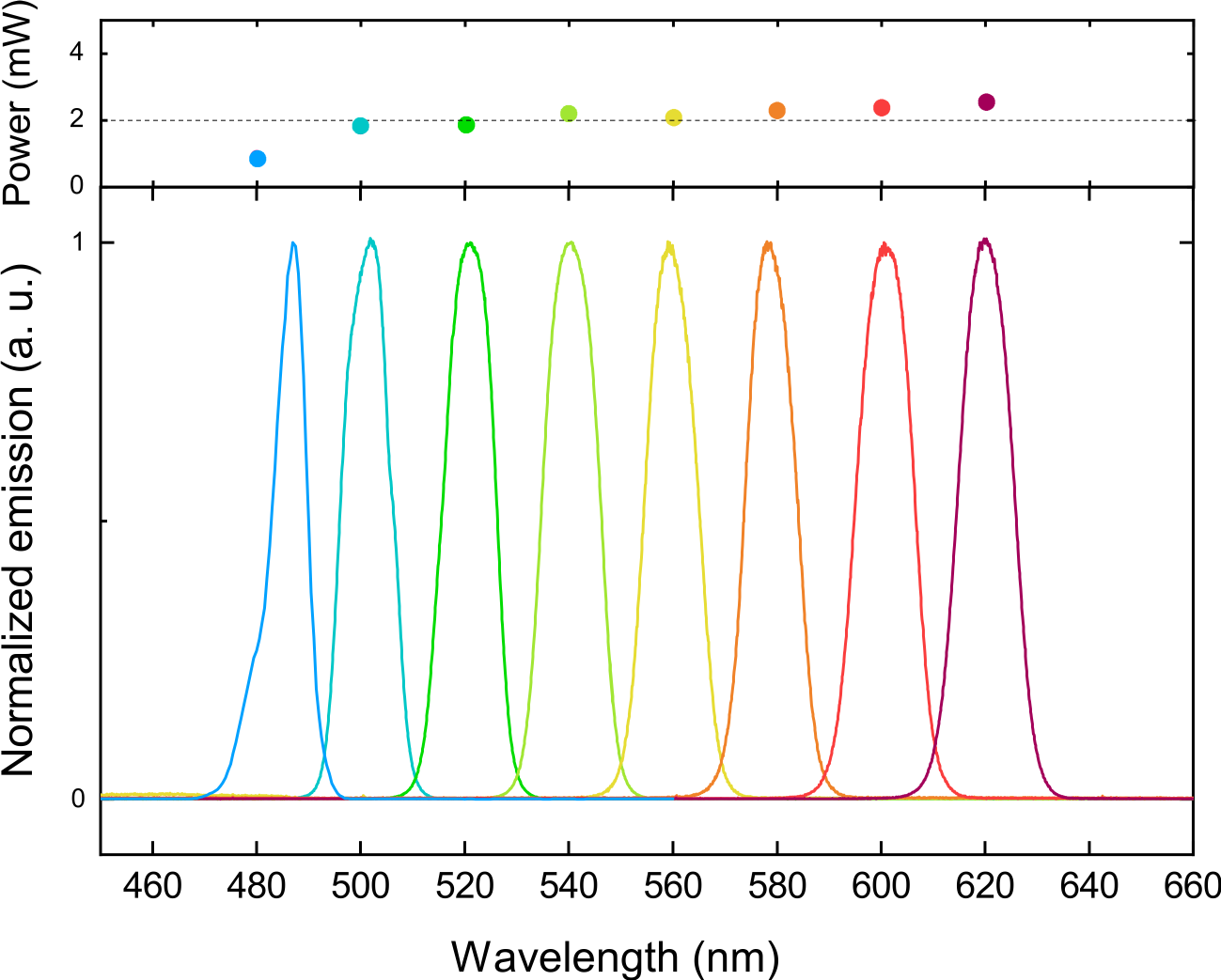}
\caption{Spectrum of the polarizing pulses (bottom) and corresponding polarizing power (top) at each wavelength used. Linewidth is 10 nm, except for the line at 480 nm, having bandwidth 20 nm. 
}
\label{WL}
\end{center}
\end{figure}

The experiments are carried out by first calibrating the NIR trap at each designated distance from each surface. Thus, we measure and analyse the total force on the trapped particle ($F_z$) arising under the illumination of the (chopped) polarizing laser beam (Fig. \ref{Fig1}\textbf{C}) as a function of wavelength, $\lambda_{\rm PB}$, and laser power. In Fig. \ref{WL}, the spectrum of the polarizing pulses (bottom) and the power (top) at each wavelength used is shown. Bandwidth is 10 nm, except for the line at 480 nm, with bandwidth 20 nm to increase the beam power. Due to reduced transmission of the SuperVaria monochromator for $\lambda<$490 nm, this line has an asymmetric shape.

%Fig.WL

\section*{Supplementary text}
\subsection*{S1 Power spectrum density analysis of a trapped particle fluctuations}

\begin{figure}
\begin{center}
\includegraphics[width=0.8\textwidth]{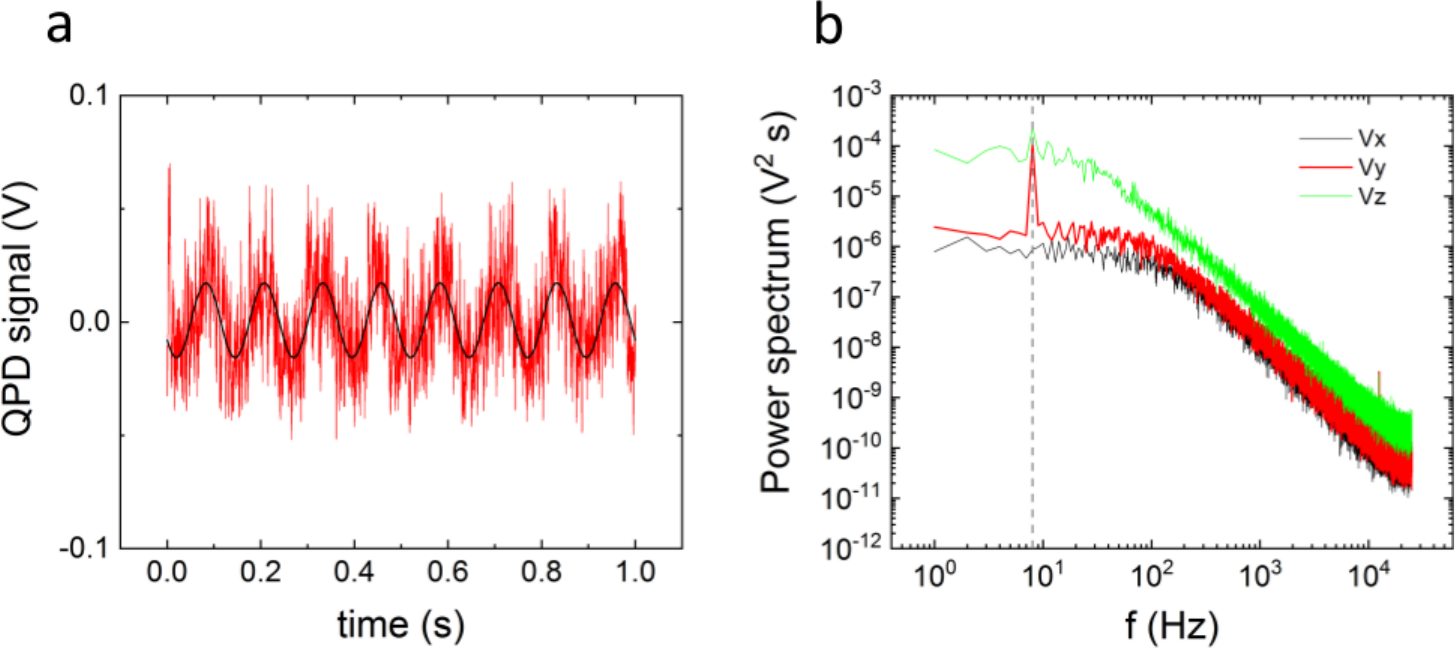}
\caption{ a) Drag force method. A trapped particle is subjected to a sinusoidal oscillation of known amplitude ($A_s$) in a direction ($y$) parallel to the surface (a). The peak (b) in power spectral density of the corresponding signal is used to calculate the diffusion coefficient $D_{\|}$ as a function of the distance $d$ of the beam focus from the surface.}
\label{S1}
\end{center}
\end{figure}

In overdamped regime, the dynamics of a trapped spherical particle follows the Langevin equation \cite{Jones2015}:

%\usetagform{S}
%if you want to remove the S, use the command \usetagform{default}
\begin{equation}
\frac{dx(t)}{dt}+ \frac{k_{x}}{\gamma}  x(t)=\sqrt{2D}W_{x}(t)
\label{Lang_eq}
\end{equation}

\noindent where for simplicity only one of the spatial directions is considered. In this equation, $k_{x}$ is the trap stiffness in the $x$ direction, $\gamma=6\pi\eta R$ is the particle friction coefficient, $R$ is the sphere radius, $\eta$ is the medium viscosity and $D=k_B T/\gamma$ is the diffusion coefficient. $W_{x}(t)$ is a white noise, representing a stochastic force with $\langle W(t) \rangle=0$, $\langle W(t)^2 \rangle$=1 for each $t$ and $W(t_1)$ independent from  $W(t_2)$ for  $t_1$ $\neq$ $t_2$ \cite{Jones2015}. Fourier transforming Eq. S\ref{Lang_eq} and calculating the square modulus of both sides we obtain:

\begin{equation}
P^B(f)=2\vert \hat{X}(f)\vert^2=\frac{D}{\pi^2 (f_{c}^2 +f^2)}
\label{psd}
\end{equation}

\noindent where the corner frequency $f_c=\frac{k_{x}}{2\pi \gamma}$ is introduced. Eq. S\ref{psd} is the \textit{power spectral density} (PSD) of the Brownian fluctuations of the trapped particle. The fit with a Lorentzian of this power spectrum gives both the corner frequency $f_c$ and the amplitude $D/\pi^2$ of the spectrum. However, in our measurements we use a position sensitive detector, giving the tracking signal in Volt (V) units. Thus, the experimentally obtained power spectrum is in $\mathrm{V^2 \cdot s}$ units, and its amplitude related to the diffusion coefficient $D$ by means of a calibration factor $\beta^2 =\frac{D}{D^{exp}}$. If the theoretical diffusion coefficient $D$ is known, namely, if the radius of the particle and the viscosity of the medium (and, thus, its temperature) are known, by fitting the experimental power spectrum the $\beta$ calibration factor can be calculated. Moreover, by using the corner frequency also the trap spring constants is obtained, giving a full calibration of the trap \cite{Jones2015,Berg2004}.

When the diffusion coefficient is not exactly known, a different approach must be used \cite{Tolic-Norrelike}. In this case, the sample chamber is subjected to a known sinusoidal oscillation:

\begin{equation}
x(t)=A_{s}\  \mathrm{sin}(2 \pi f_s t)
\label{drag_mov}
\end{equation}

\noindent where $A_s$, in length units, is the known amplitude of oscillation and $f_s$ is the oscillation frequency. The trapped particle will thus be subjected to a similarly oscillating drag force (Fig. \ref{S1}a)

\begin{equation}
F_{drag}(t)=-\gamma v_s =-\gamma 2 \pi f_s A_{s} \mathrm{cos}(2 \pi f_s t)
\label{drag_force}
\end{equation}

\noindent where $v_s$ is the stage velocity. In this condition, the Langevin equation becomes:
\begin{equation}
\left(\frac{dx(t)}{dt}-v_s\right)+\frac{k_{x}}{\gamma} x(t)=\sqrt{2D}W_{x}(t)
\label{Lang_eq_dragforce}
\end{equation}

and the PSD will be:

\begin{equation}
P(f)=\frac{D}{\pi^2 (f_{c}^2+f^2)}+\frac{A_s^2}{2(1+\frac{f_{c}^2}{f_s^2})}\delta (f-f_s)
\label{psd_modulation}
\end{equation}

This power spectrum has a Lorentzian profile to which a delta function at the stage frequency of oscillation (see Fig. \ref{S1} b, red curve) is overlapped. The amplitude of this peak can be used to obtain the calibration factor $\beta$ in units of m/V

\begin{equation}
\beta=\sqrt{\frac{P_{th}}{P_{ex}}}
\label{beta}
\end{equation}

\noindent where $P_{th}$ is the theoretical value for the peak at the stage frequency

\begin{equation}
P_{th}=\frac{A_s^2}{2\left(1+\frac{f_{c}^2}{f_s^2}\right)}
\label{P_th}
\end{equation}

\noindent which is in $m^2$ units, whereas the $P_{ex}$, in $V^2$ units, is
  
\begin{equation}
P_{ex}=\left[ P(f_s)-P^B(f_s)\right]\Delta f
\label{P_ex}
\end{equation}

\noindent where $\Delta f=\frac{1}{t_{m}}$ is the frequency resolution, $t_{m}$ is the measurement time, $P^B$ is the value of the power spectrum, at the stage frequency, due to the standard Brownian motion (Eq. S\ref{psd}). In this way, we can calculate $\beta$ to calibrate the fitted amplitude of the power spectrum and obtain the diffusion coefficient. Finally, if the temperature is known, the drag coefficient $\gamma=\frac{k_B T}{D}$ can be calculated and, thus, the trap spring constant $k_x=2 \pi f_{c} \gamma$ is easily obtained \cite{Jones2015,Tolic-Norrelike}. In principle, this method can be applied to each direction of the space, giving the full 3D calibration of the trap.

%\newpage

\subsection*{S2 Hydrodynamics of a sphere close to a surface and NIR trap calibration}
Close to a surface, the motion of a sphere is hindered with respect free motion in bulk. As a consequence, the diffusion coefficient in direction  
parallel to the surface, $D_{\|}$,  is different from that in perpendicular direction $D_{\bot}$, and both depend on the distance $\xi$ between the center of the particle and the surface itself. $D_{\|} (\xi)$ is given by the Faxen's law \cite{Faxen1922widerstand}:

\begin{equation}
D_{\|}(\xi)=D \cdot \left(1-\frac{9R}{16\xi}+\frac{R^3}{8\xi^3}-\frac{45R^4}{256\xi^4}-\frac{R^5}{16\xi^5} \right)
\label{D_par}
\end{equation}

\noindent while for $D_{\bot}(\xi)$ we use 

\begin{equation}
D_{\bot}(\xi)=D \cdot \left(1-\frac{9R}{8\xi}+\frac{R^3}{2\xi^3}-\frac{57R^4}{100\xi^4}+\frac{R^5}{5\xi^5}+\frac{7R^{11}}{200\xi^{11}}-\frac{R^{12}}{25\xi^{12}}\right)
\label{D_perp}
\end{equation}

\noindent which is an approximation \cite{SchaefferLANGM07} 
deviating less than 0.3\% from the exact infinite sum formula given by other authors \cite{Brenner1961slow}. Obviously, for $\xi \rightarrow \infty$, $D_{\|}=D_{\bot}=D=\frac{k_B T}{6\pi\eta R}$.

In our measurements, we have chosen to fully calibrate the trap, at decreasing distances from the surface, with the drag force method discussed in the previous section to reconstruct the Faxen's law. This approach has been used by Tolic-Norrelike et al. \cite{Tolic-Norrelike}, which modified Eq. S\ref{D_par} taking in account that the real distance from the surface is the difference between the position $d$, as controlled by the piezostage, and the absolute position $h_s$ of the surface. Moreover, also the focal shift $\delta$ due to aberrations connected to the objective used must be taken in account in the evaluation of the right distance from the surface. Thus, $\xi=\left(d-h_s\right)\delta$ and Eq. S\ref{D_par} must be corrected accordingly:

\begin{equation}
\frac{D_{\|}(d)}{D} = K \left(1-\frac{9R}{16\left(d-h_s\right)\delta}+\frac{R^3}{8\Big(\left(d-h_s\right)\delta\Big)^3}-\frac{45R^4}{256\Big(\left(d-h_s\right)\delta\Big)^4}-\frac{R^5}{16\Big(\left(d-h_s\right)\delta\Big)^5} \right)
\label{Dpar_corretta}
\end{equation}

In our case, as we use an oil-immersion objective with NA=1.3, the focal shift is $\delta=0.81$ \cite{SchaefferLANGM07}. 

\begin{figure}
\includegraphics[width=\textwidth]{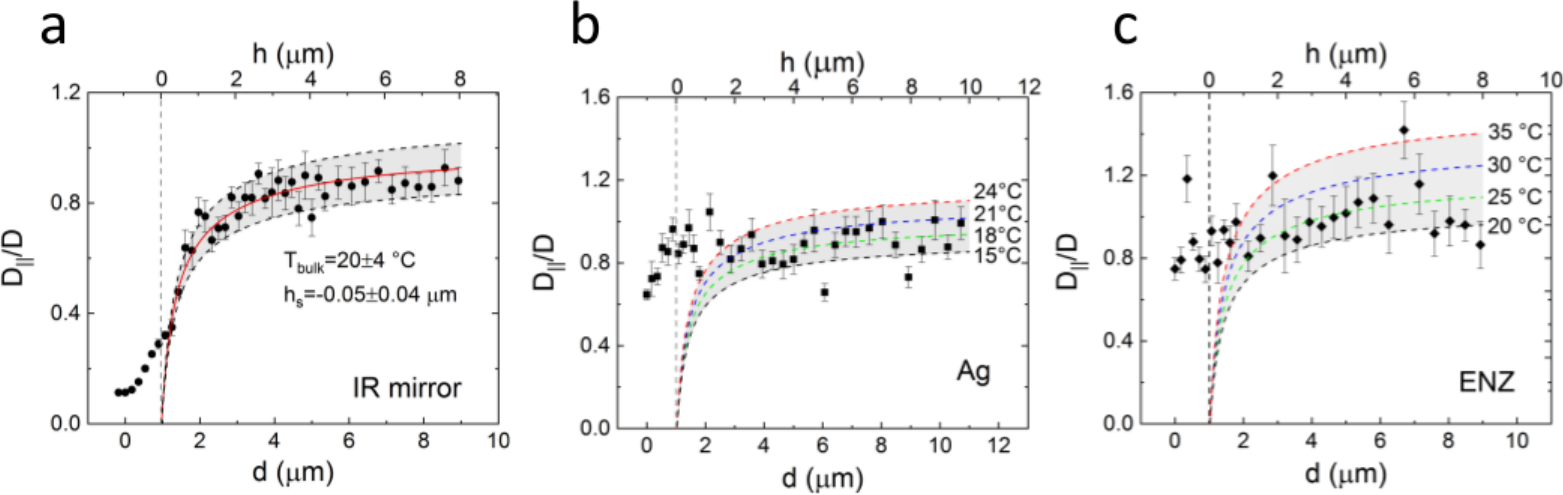}
\caption{ a) Fit of $D_{\|}(d)$ following the Faxen law. The data have been normalized to the theoretical diffusion coefficient $D$ of a 2 $\mu$m sphere at the ambient temperature $T_{amb}$. The fit allows to evaluate the absolute surface position $h_s$ and the sample bulk temperature $T_{bulk}$. The gray shaded region represents the variability on the estimation of the medium bulk temperature. In case of a pure reflective surface, as for a IR mirror (a), $T_{bulk} \sim T_{amb}$ (that is, $D_{\|}/D \rightarrow $ 1) and $h_s$ is consistent with the estimation of the absolute surface position made by visual inspection (in general, $h_s = 0\pm 1$ $\mu$m). In front of Ag surface (b) and ENZ surface (c), a deviation from the Faxen law is observed, giving in both cases larger diffusion coefficients close to the surfaces and, in case of ENZ surface, $D_{\|}/D >1$. Thus, the estimation of the sample temperature, for each distance $d$ from the surface, has been done by considering, at fixed $h_s=0\,$ $\mu$m, a theoretical Faxen law (see text) close to the measured diffusion coefficient at that distance. With this procedure, we estimate a maximum 5\% error in the value of the temperature in front of the ENZ surface.}
\label{S2}
\end{figure}
In principle, a similar procedure could be also used to calibrate the trap in the axial direction; however, in this case there would be some limitations \cite{SchaefferLANGM07}. Differently, in our measurements we indirectly calibrated the trap in the axial direction by using the values of the surface absolute position $h_s$ and of the bulk diffusion coefficient $D$, the latter at the bulk temperature evaluated with the lateral calibration procedure. These parameters are used in the equation:

\begin{align}
D_{\bot}(d)= D \cdot \Bigg(1-\frac{9R}{8\left(d-h_s\right)\delta}+\frac{R^3}{2\Big(\left(d-h_s\right)\delta\Big)^3}-\frac{57R^4}{100\Big(\left(d-h_s\right)\delta\Big)^4}+ \nonumber
\\
+\frac{R^5}{5\Big(\left(d-h_s\right)\delta\Big)^5}+\frac{7R^{11}}{200\Big(\left(d-h_s\right)\delta\Big)^{11}}-\frac{R^{12}}{25\Big(\left(d-h_s\right)\delta\Big)^{12}} \Bigg)
\label{D_perp_mod}
\end{align}

\noindent From the fitted amplitude and frequency corner of the axial PSD we can also obtain the calibration factor in axial direction $\beta_z$

\begin{equation}
\beta_z(d)=\sqrt{\frac{D_{\bot}(d)}{D_{\bot}^{volts}(d)}}
\label{beta_zeta}
\end{equation}

\noindent and the axial trap spring constant 

\begin{equation}
k_z(d)=2 \pi f_{c,z} \frac{k_B T}{D_{\bot}(d)}
\end{equation}

If the ambient temperature is known, and thus an estimation of the medium bulk diffusion coefficient $D$ can be given, the fit with Eq. S\ref{Dpar_corretta} of the $D_{\|}(d)$ values obtained with the drag force method gives the absolute position $h_s$ and the parameter $K$. In the limit $d \rightarrow \infty$, $K$ should tend to unity; if this condition is not met and, for instance, the value of $K$ is higher, the initial estimation of the bulk temperature becomes inaccurate, leading to an underestimation. Consequently, the actual temperature of the medium is higher than initially estimated.%

In our measurements, a 8 Hz sinusoidal oscillation in the $y$ direction was applied to the piezostage  (see Fig. \ref{S1}a). The amplitude of this oscillation is comprised between 120 nm and 180 nm, depending on the surface studied. We calculated Power Spectral Density (PSD, Fig. \ref{S1}b) of the signal when the trapped particle approaches the surface, and by the fit of the spectrum in the $y$ direction we extracted the calibration factors (Eq. S\ref{beta}) and the diffusion coefficients $D_{\|}(d)$ for each distance from the surface. Then, $D_{\|}(d)$ has been fitted with Eq. S\ref{Dpar_corretta} to reconstruct the Faxen's law and obtain an estimation of the absolute position of the surface and of the bulk temperature of the medium, the latter by means of the parameter $K$. In Fig. \ref{S2}a, it is shown that in front of a IR reflective mirror the data can be fitted with a curve to which correspond a medium bulk temperature $T_{bulk}$ of approximately 20 $\pm$ 4 $^\circ$C and an absolute surface position of approximately $h_s$=-0.05 $\pm$ 0.04 $\mu$m. %

In general, in front of dielectric surfaces (glass or a IR mirror), Faxen's law for the $D_{\|}$ is quite well followed (red curve in Fig. \ref{S2}a), even if a high dispersion in the data can be observed, due to the %possible% 
low intensity of the backscattered signal (in front of glass surface) and/or to the low trapping power used (at maximum 40 mW at the objective, against much higher values used in these type of measurements by other authors \cite{SchaefferLANGM07}). However, in front of metallic (Ag) or ENZ surfaces, the adherence to Faxen's law diminishes significantly at small distances $d$ (see Fig. \ref{S2}b and c). 

In these cases we used very low NIR trapping powers (approximately 4 mW in front of Ag surface and 2 mW in front of ENZ surface). However, even at the smallest laser power compatible with particle trapping, Faxen's law is not well followed, allowing only the evaluation of a bulk temperature, but not of the $h_s$ parameter. Therefore, we set it to $h_s$=0 $\mu$m, ensuring precision by meticulously focusing the laser beam onto the surface at the onset of the measurements. We then verified that this reference point was maintained consistently throughout all measurements conducted to reconstruct Faxen’s law. In Fig.  \ref{S2}b we see that the bulk temperature of Ag surface is compatible with the ambient temperature (approximately 18 $^\circ$C) measured before the measurements; heating effects, increasing the diffusion coefficient, are seen only close to the surface. In front of ENZ surface, we find a bulk temperature of 27.5 $\pm$ 7.5 $^\circ$C, larger than ambient temperature and with a higher level of uncertainty (gray dashed region), and again large diffusion coefficients close to the surface (Fig. \ref{S2}c). 
Aiming at calculating $F_z$, the estimation of the medium temperature, at each distance $d$ from the Ag and ENZ surfaces, has been done by considering, at fixed $h_s=0\,$ $\mu$m, a theoretical Faxen law close to the measured diffusion coefficient at that distance. 
With this procedure, we estimate a maximum 5\% error in the value of the temperature.

\subsection*{S3 Pulsed force on a trapped particle and power spectral density}
%Fig.4
\begin{figure}
\begin{center}
\includegraphics[width=0.6\textwidth]{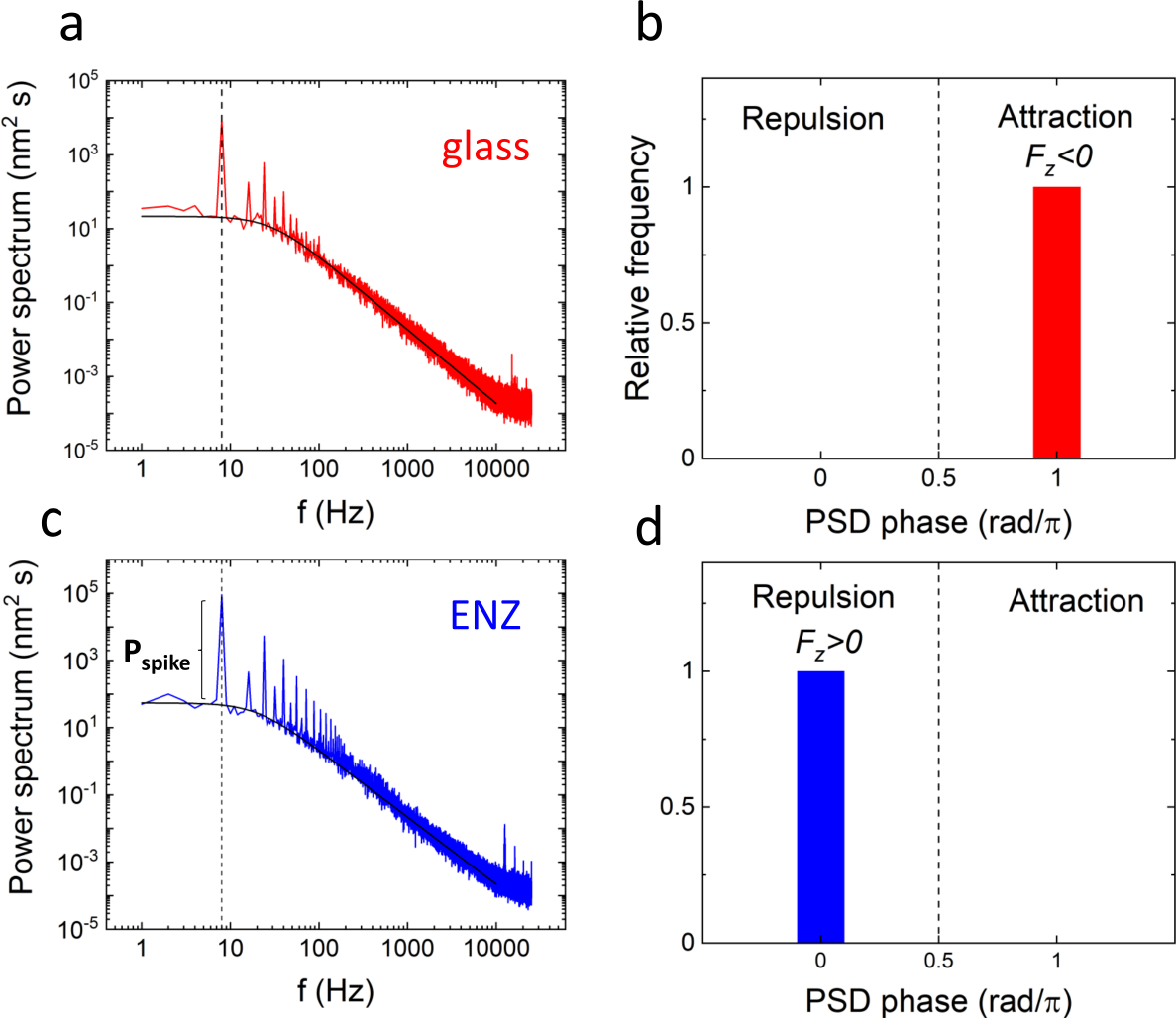}
\caption{(\textbf{a, c}) Calibrated power spectral density (PSD) of the axial signal for a bead in front of glass (a)  or ENZ (c) surfaces, under polarizing beam illumination at 8 Hz. Each spectrum is an average over ten power spectra, each one corresponding to 1 s observation time, at 50 KHz sampling rate. (\textbf{b,d}) The value of the PSD phase allows the recognition of the attractive (b) or repulsive (d) character of the total force.}
\label{FigS3}
\end{center}
\end{figure}
We consider a periodic perturbation on the trapped particle fluctuations due to a pulsed axial external force \cite{Zensen_APL2016}. The external perturbation of the trapped particle thermal fluctuations is modelled in terms of the Heaviside step function $H$: 

\begin{equation}
F_z (t)={ e}^{i\Phi}F_z\cdot \sum_{0}^{k=N-1} \left[ H(t-kt_p)-H(t-kt_p -\frac{t_p}{2})\right] 
\label{Heaviside}
\end{equation} 

\noindent where $F_z$ is the amplitude of the applied force, $\Phi$ is a global phase describing the attractive, $\Phi=\pi$, or repulsive, $\Phi=0$, nature of the force, $N=\frac{t_m}{t_p}$ is the number of pulse repetitions, $t_m$ is the total measurement time and $t_p$ is the period of the applied force. Adding this perturbation to the Langevin equation (Eq. S\ref{Lang_eq}) and taking the Fourier transform, the resulting power spectral density is \cite{Zensen_APL2016}:

\begin{equation}
P(f)=\frac{D^2}{(2\pi k_B T)^2}\frac{1}{f_{c}^2+f^2}\left[ 4 \frac{(k_B T)^2}{D} + \frac{F_{z}^2}{4\pi^2 f^2 t_m} \cdot \left(\frac{\mathrm{sin}\pi f t_m}{cos \frac{\pi}{2}f t_P}\right)^2\right] 
\label{Fs_completa}
\end{equation}

\noindent The first term in this sum is the standard diffusive motion of the particle whereas the second term is the due to the periodic pushing force that, in the limit $f\rightarrow 1/t_p$ gives:

\begin{equation}
P_{spike}=\frac{1}{4\pi^4}\left(\frac{D}{k_B T}\right)^2\frac{t_m}{\left(f_{c}^2+f_{p}^2\right)}F_{z}^2
\label{S_fchopper}
\end{equation}

\noindent where $P_{spike}$ (see Fig.\ref{FigS3}c) is the power at the light pulse frequency $f_p=1/t_p$ minus the Brownian contribution in the power spectrum at the same frequency. From this equation, the amplitude of the external axial force is obtained as:

\begin{equation}
F_z=2 \pi^2 \frac{k_B T}{D} \sqrt{P_{spike} \frac{1}{t_m} \left(f_{c}^2+f_{p}^2 \right)}
\label{Fscatt}
\end{equation} 
where Eq. S\ref{D_perp_mod} is used to calculate, at each distance from the surface, the diffusion coefficient.

Equation S19 provides the amplitude of the force $F_z$, yet it lacks information regarding its direction, whether it is attractive or repulsive.
However, the phase of the PSD at $f$=0 (as shown in Fig. \ref{FigS3}b and d), is linked to the signal phase, $\Phi$. Specifically, if the tracking signal exhibits a negative shift (illustrated by the red curve in Fig.2 a), the phase at $f$=0 is $\Phi=\pi$ (Fig. \ref{FigS3}b), indicating an attractive force.  Conversely, if the tracking signal displays a positive shift (as shown by the blue curve in Fig.2 a), the phase at $f$=0 is $\Phi=0$ (Fig. \ref{FigS3} d),  signifying a repulsive force.%   

It is worth noting that the PSD method is extremely accurate when measuring a small external force and its sign. For example, when the shift of the external force is within the particle positional fluctuations the PSD analysis can easily distinguish a small peak on the wide Lorentzian thermal profile and the phase at $f=0$ provides a simple way to distinguish between an attractive or repulsive force.% 

\subsection*{S4 Light-driven repulsive force as a function of ENZ excitation power: nonlinear behaviour, thermophoresis, and particle velocity}
%Fig.5
\begin{figure}
\centering
\includegraphics[width=0.6\textwidth,angle=270]{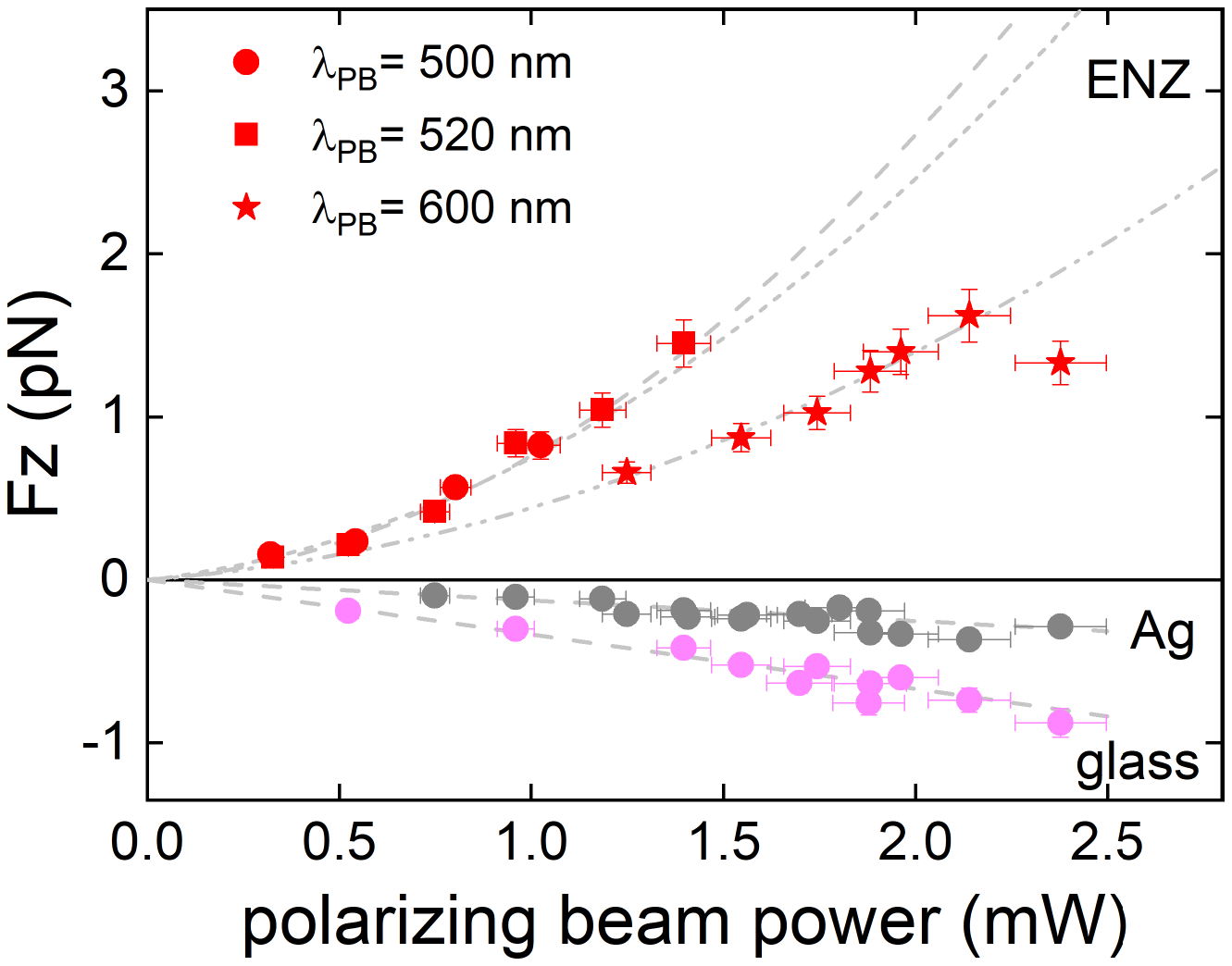}
\caption{Optical force at fixed PB wavelengths (500 nm, 520 nm, 600 nm) as a function of the pulse power. While in front of glass and Ag the optical force is, as expected, proportional to the pulse power, in front of ENZ a second order polynomial is needed to fit the data. Height is 0.8 microns for ENZ and Ag, 1.3 microns for glass. 
}
\label{Fig5}
\end{figure}
Aiming at understanding the origin of the behaviour of the axial force in front of ENZ surface, we have studied the dependence of the force on the power of the laser pulses (Fig. \ref{Fig5}). In case of ENZ, we fixed the edge-to-edge distance ($h$=0.8 $\mu$m from the surface) and considered three different wavelengths, just on the maximum of the force (500 nm), close to the maximum (520 nm) and far away from the maximum (600 nm). It is interesting to note that, at 500 nm, it was not possible to increase power above approximately 1 mW because the repulsive force was so high to expel the particle from the optical trap. Higher powers can be studied with 520 nm, but also in this case the bead is expelled from the trap. In all cases, we observe that the axial force exhibits a second-order dependence on power at each wavelength. Conversely, a similar analysis conducted in front of glass and Ag surfaces reveals a linear relationship between the force and power.

%\subsection*{S7 Particle velocity under ENZ excitation}
%Fig.4
\begin{figure}
\begin{center}
\includegraphics[width=0.5\textwidth]{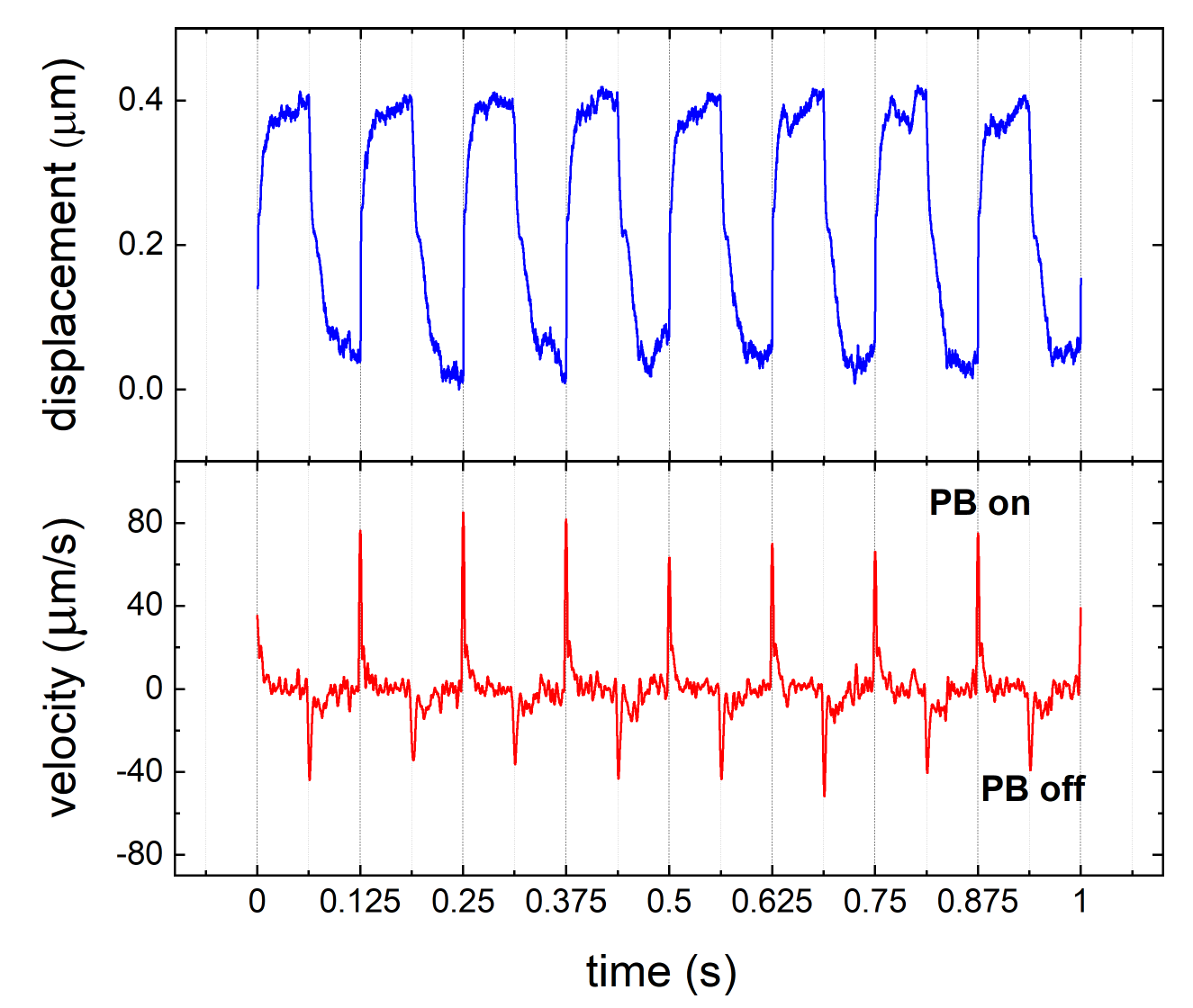}
\caption{Calibrated signals (blue curve) registered in front of ENZ surface under polarizing beam pulses and (red curve) correspondent its first order derivative. The spikes at positive (or negative) values correspond to the values of the particle velocity away from (positive) or towards (negative) the surface following the PB on/off illumination.}
\label{FigS4}
\end{center}
\end{figure}

Particle dynamics can be further analyzed by measuring the velocity under the pulsed ENZ excitation (PB on) and during restoration of the particle equilibrium position (PB off). In Fig. \ref{FigS4} the signal (blue curve) obtained with the 2 $\mu m$ bead trapped in front of the 547 nm ENZ surface shows a square wave modulation. The rising part of the signal corresponds to the repulsive motion away from the surface with the PB on, while the descending part corresponds to the particle recovering its equilibrium position in the NIR trap after the PB is switched off. Correspondingly, in the bottom panel the first derivative of this signal is shown (red curve), with positive (negative) spikes associated to particle moving away from/towards the surface. Thus, the particle is repelled from the ENZ surface %by the light-driven ENZ force%
with a velocity in the range 60-80 $\mu m$/s, while it recovers its NIR trap equilibrium position with a velocity of about 40 $\mu m$/s. This different repulsive/restoring dynamics is related to the different forces acting on the particle when the polarizing beam is on/off. When the PB is turned on, the particle is pushed away at a speed that is almost double with respect to the speed of the restoring motion after the PB is switched off. 

\begin{figure}
\begin{center}
\includegraphics[width=0.6\textwidth]{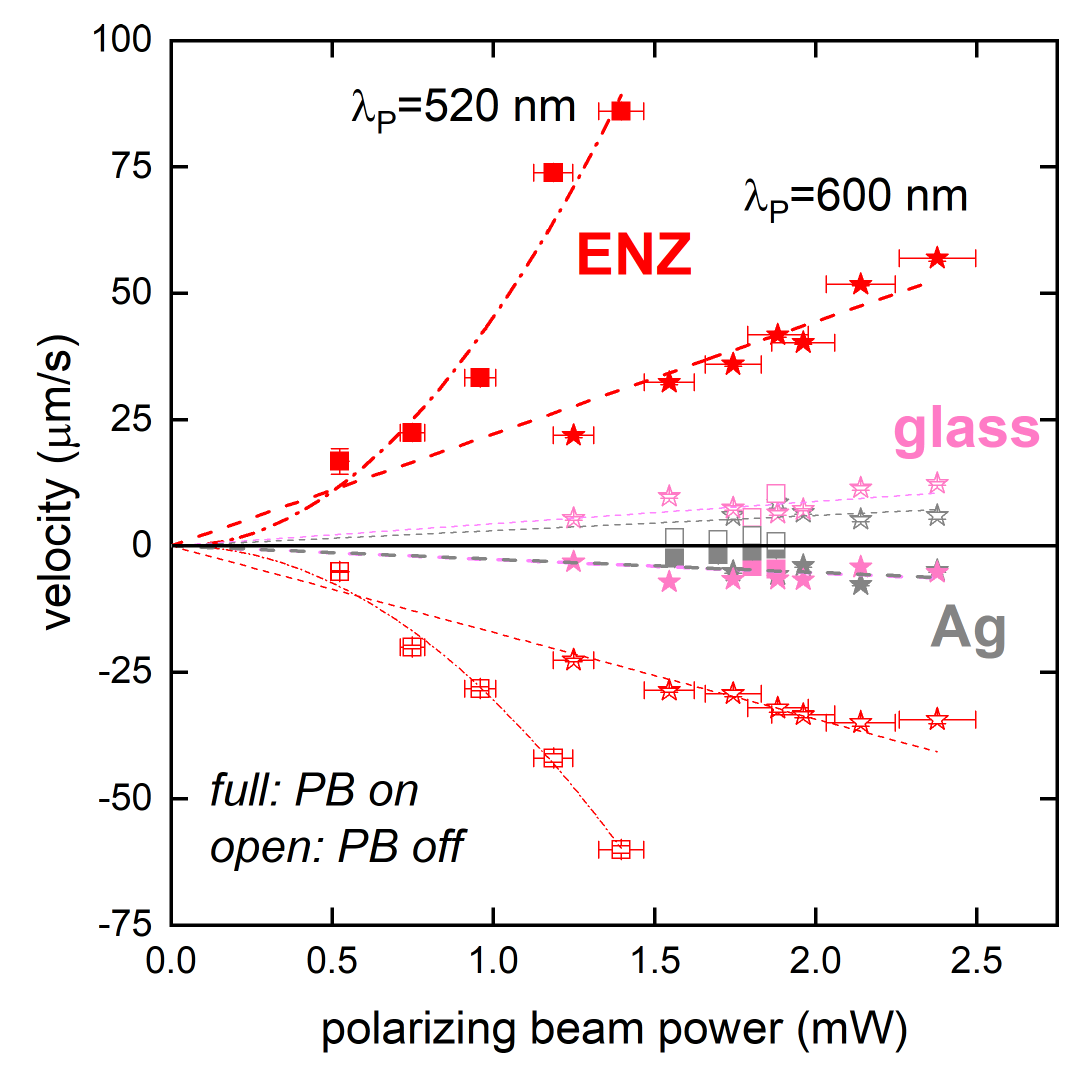}
\caption{Particle velocity as a function of increasing power of the PB pulses and at two different wavelengths (squares: $\lambda_p$=520 nm; stars: $\lambda_p$=600 nm). Full points correspond to PB switching on, while open points correspond to PB switching off. Data measured in front of ENZ (red), Ag (gray) and glass (pink) surfaces are shown.}
\label{FigS5}
\end{center}
\end{figure}

In Fig. \ref{FigS5} the particle instant velocity with/without PB illumination (full/open symbols) at $\lambda_p$=520 nm and  $\lambda_p$=600 nm in front of ENZ (red), Ag (gray) and glass (pink) surfaces is shown at increasing PB illumination power. A different dependence on laser power close to and far from the ENZ wavelength (520 nm and 600 nm, respectively) is observed in the case of the ENZ surface, while a weakly linear dependence is observed for Ag and glass surfaces. Since optical forces are linearly dependent on power, we expect that the particle displacement induced by the laser pulses has a linear dependence on the pulse power. This is consistent with observations at 600 nm (red stars in Fig. \ref{FigS5}) and for glass and Ag surfaces. The nonlinear dependence of the velocity on the pulse power at 520 nm suggests that close to the ENZ wavelength other forces add to the radiation pressure, driving the system out of the linear regime.  

%Fig.S5

\subsection*{S5 Absorbed power spectrum}

\begin{figure}[t]
    \centering\includegraphics[width=0.6\textwidth]{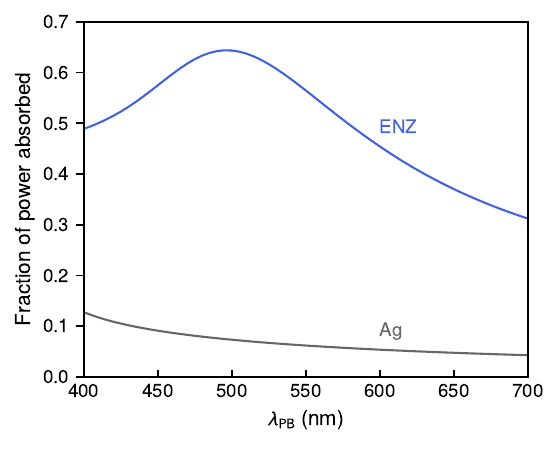}
    \caption{Fraction of the incident power absorbed as a function of polarizing beam wavelength $\lambda_{\rm PB}$ for two different surface configurations: a 547 nm ENZ surface (blue curve) versus an Ag surface (gray curve). The calculation is done using a geometrical optics approximation assuming the incident polarizing beam has a Gaussian profile with waist $w_0 = 2.8$ $\mu$m. Centered on the beam axis is a 1 $\mu$m radius polystyrene particle at edge-to-edge height $h=0.8$ $\mu$m above the surface.}\label{FigAbsorption}
\end{figure}

Since the possible thermophoretic contribution to the total force is related to the power absorbed by the substrate from the incident polarizing beam, we theoretically calculated the wavelength dependence of this absorption for two different substrate types. Fig.~\ref{FigAbsorption} shows the fraction of incident power absorbed as a function of polarizing beam wavelength $\lambda_{\rm PB}$ in the range 400-700 nm. The blue curve corresponds to the 547 nm ENZ substrate, and the gray curve to an Ag substrate. In both cases a geometrical optics approximation was employed, using parameters that mimicked the experimental setup: the polarizing beam had a Gaussian profile with waist $w_0 = 2.8$ $\mu$m, and a 1 $\mu$m radius polystyrene particle was placed along the beam axis at edge-to-edge height $h=0.8$ $\mu$m above the surface. The ENZ substrate exhibits much larger power absorption compared to the Ag one (which mainly reflects the beam in this wavelength range). The ENZ absorbed power spectrum peaks around 495 nm, similar to location of the peak in total repulsive force seen in main text Fig.~3. However the changes in absorption in the ENZ wavelength region are far more gradual than the dramatic increases in the experimentally measured total force: we see increases of around $30\%$ in absorbed power going from 580 nm to 500 nm, compared to more than $300\%$ in the total force over the same wavelength range for particles at heights $h = 0.8$ $\mu$m and $1.7$ $\mu$m. This suggests that thermophoretic forces, while likely a contributing factor, cannot fully explain the behavior observed in Fig.~3.

\subsection*{S6 Additional ENZ sample with a titania-silver layered structure}
\begin{figure}
\begin{center}
\includegraphics[width=0.6\textwidth]{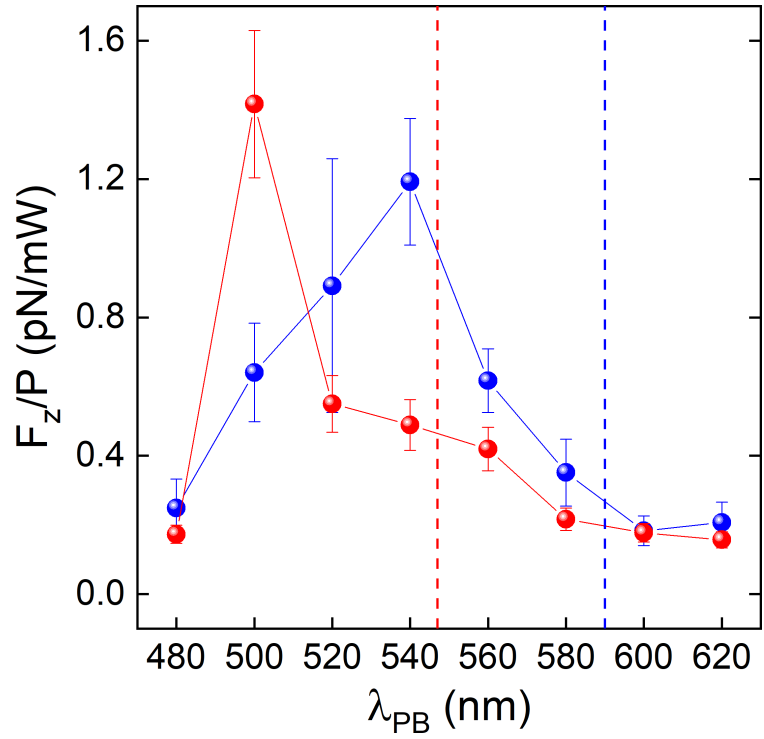}
\caption{Normalized axial force in front of two different ENZ surfaces with distinct ENZ frequency ranges exhibits similar behaviors, confirming the observed levitating response. The data illustrates a blue shift of the peak maximum in comparison to nominal ENZ wavelengths (indicated by red and blue dashed lines). 
Notably, the force measurements adhere to the absorption trend, shown in Fig. S8, outside the ENZ region but diverge from it within the ENZ region where the repulsion force shows a resonant sevenfold increase.}
\label{FigENZ}
\end{center}
\end{figure}
We have carried out force measurements as a function of the wavelength of the polarizing beam in front of an ENZ sample at 590 nm. The 590 nm ENZ sample consisted of the following structure, from top to bottom: TiO$_2$ (3 nm) / Ag (19 nm) / [8 trilayers:  Ge (1 nm) / TiO$_2$ (40 nm) / Ag (19 nm)] / Ti (3 nm). The stacks were deposited on a glass substrate (Corning Inc.) with electron-beam evaporation for Ge (0.5 $\text{\AA}  / \text{s}$) and thermal evaporation for Ag (0.5 $\text{\AA} / \text{s}$).  Ti and TiO$_2$ thin films were deposited by dc magnetron sputtering on top of the ENZ multilayer. A metallic Ti layer was first grown in an Ar discharge to promote adhesion, and TiO$_2$ was subsequently obtained by reactive sputtering of Ti in a mixed Ar/O$_2$ atmosphere. Also for this sample the materials used in the fabrication were purchased from Kurt J. Lesker and to determine thicknesses and refractive indices of the layers in each stack, ellipsometric measurements in an air superstrate were carried out using a variable-angle, high-resolution spectroscopic ellipsometer (J. A. Woollam Co., Inc, V-VASE) for incident angles $45^\circ$, $50^\circ$, $55^\circ$ and wavelength range $300 - 1000$ nm.  Once the parameters of the stack were established via best-fits to the ellipsometric data, the ENZ wavelength was determined via the effective medium theory described in Ref.~[Ref. 33]\cite{kiasat2023epsilon}.

The force measurements results are shown in Fig. \ref{FigENZ}. Also for this sample we observe a maximum of the force at a wavelength close to the ENZ wavelength (Fig \ref{FigENZ}, blue curve). The comparison between this curve and the one obtained an the 547 nm ENZ sample shows that both maxima are blue-shifted with respect to the nominal ENZ wavelength with about a sevenfold increase in the net repulsive force in the ENZ region with respect to the force away from the ENZ region.

\bibliographystyle{ieeetr} 
\bibliography{biblioENZ}

\newpage

\end{document}